

\input amstex
\documentstyle{amsppt}

\loadbold
\CenteredTagsOnSplits

\topmatter
\title
Multidimensional dynamical systems\\ accepting the normal shift.
\endtitle
\author
Boldin A.Yu. and Sharipov R.A.
\endauthor
\address
Department of Mathematics,
Bashkir State University, Frunze str. 32, 450074 Ufa,
Russia
\endaddress
\email
root\@bgua.bashkiria.su
\endemail
\date
April 12, 1993.
\enddate
\abstract
The dynamical systems of the form $\ddot\bold r=\bold F(\bold r,
\dot\bold r)$ in $\Bbb R^n$ accepting the normal shift are
considered. The concept of weak normality for them is
introduced. The partial differential equations for the force
field $\bold F(\bold r,\dot\bold r)$ of the dynamical systems
with weak and complete normality are derived.
\endabstract
\endtopmatter
\document
\head
1. Introduction
\endhead
     The concept of dynamical system accepting the normal shift
was introduced in \cite{1} as a result of generalization of the
classical geometrical Bonnet transformation (or normal shift)
for the case of dynamical systems. In \cite{1} (see also
\cite{2} and \cite{3}) the dynamical systems in $\Bbb R^2$ are
studied (see \cite{1}, \cite{2} and \cite{3} for detailed
reference list). In present paper we generalize the results of
\cite{1} for the multidimensional case and considser some
peculiarities absent in 2-dimensional case. These results are
declared in \cite{4}. They also form the section 5 in preprint
\cite{3}. \par
     Let's consider the dynamical system describing the
trajectories $\bold r=\bold r(t)$ in $\Bbb R^n$ particles with
unit mass in the force field $\bold F(\bold r,\dot\bold r)$
$$
\ddot\bold r=\bold F(\bold r,\dot\bold r)\tag1.1
$$
We shall use the trajectories of \thetag{1.1} for transforming
the submanifolds of $\Bbb R^n$. Let's consider the set of
particles each of which is starting at $t=0$ from some point
$P$ on some hypersurface $S\subset\Bbb R^n$ in the direction of
normal vector $\bold n(P)$ with some initial velocity $v(P)$.
In the end of time interval t these particles form another
hypersurface $S_t\subset\Bbb R^n$. This defines the one-parameter
family of hypersurfaces and the family of diffeomorphisms
$$
f_t:S\longrightarrow S_t\tag1.2
$$
Now let's recall the following two definitions from \cite{1} and
\cite{2}. \par
\definition{Definition 1} Each transformation $f=f_t$ of the
family \thetag{2.1} is called the normal shift along the
dynamical system \thetag{1.1} if each  trajectory of \thetag{1.1}
crosses each submanifold $S_t$ along its normal vector $\bold n$.
\enddefinition
\definition{Definition 2} Dynamical system \thetag{1.1} is called
the dynamical system accepting the normal shift of submanifolds
of codimension 1 if for any submanifold S of codimension 1 there
is the function $v=v(P)$ on $S$ such that the transformation
\thetag{2.1} defined by the system \thetag{1.1} and the initial
velocity function $\left|\bold v(P)\right|=v(P)$ is the
transformation of normal shift.
\enddefinition
\head
2. Normality conditions for the dynamical systems.
\endhead
     Phase space for dynamical system \thetag{1.1} is defined
by the pairs of vectors $\bold r$ and $\bold v$. In all points
of phase space where $\bold v\neq 0$ we introduce the spherical
coordinates in $\bold v$-space. Let $u^1,\dots,u^{n-1}$ be the
coordinates on the unit sphere $|\bold v|=1$ and let $u^n=v=
|\bold v|$. We define also the unit vector $\bold N$ along the
vector $\bold v$ and the derivatives of $\bold N$
$$
\bold N=\bold N(u^1,\dots,u^{n-1})\hskip 3em
\bold M_i=\frac{\partial\bold N}{\partial u^i}\tag2.1
$$
For the derivatives of the introduced vectors $\bold M_i$ one
may use the standard Weingarten formulae with the metric
connection $\vartheta^k_{ij}$ on the unit sphere
$|\bold v|=1$
$$
\frac{\partial\bold M_i}{\partial u^j}=\vartheta^k_{ij}
\bold M_k-G_{ij} \bold N \tag2.2
$$
here $G_{ij}=G_{ij}(u^1,\dots,u^{n-1})$ is the metric tensor on
unit sphere defined by scalar products $G_{ij}=\left<\bold M_i,
\bold M_j\bold M_j\right>$. In \thetag{2.2} and in what follows
coinciding upper and lower indices imply summation. Force field
for the dynamical system \thetag{1.1} may be represented by the
formula similar to that of \cite{1}
$$
\bold F = A \bold N + B^i \bold M_i\tag2.3
$$
The equation \thetag{1.1} itself then is rewritten as the
following system of differential equations with respect to
$\bold r$, $v$ and $u^i$
$$
\dot\bold r = v \bold N\hskip 3em
\dot v = A\hskip 3em
\dot u_i = v^{-1} B^i\tag2.4
$$
Let's consider the solution of \thetag{2.4} depending on some
extra parameter $s$ and introduce the following notations for
the derivatives of the coordinates in the phase space
$$
\partial_s\bold r = \tau\hskip 3em
\partial_s v = w \hskip 3em
\partial_s u^i = z^i\tag2.5
$$
Differentiating \thetag{2.4} and keeping in mind \thetag{2.1}
and \thetag{2.5} we obtain the time derivatives for
$\boldsymbol\tau$, $w$ and $z^i$
$$
\aligned
\dot{\boldsymbol\tau}&=w\bold N+v\bold M_i z^i \\
\dot w &=\frac{\partial A}{\partial r^k} \tau^k +
\frac{\partial A}{\partial v} w +
\frac{\partial A}{\partial u^k} z^k \\
\dot z^i &=-\frac{B^i w}{v^2}+\frac1v
\left(\frac{\partial B^i}{\partial r^k} \tau^k +
 \frac{\partial B^i}{\partial v} w +
 \frac{\partial B^i}{\partial u^k} z^k \right)
\endaligned
\tag2.6
$$
The equations \thetag{2.6} here are the analogs of \thetag{3.7}
from \cite{1}. In addition to \thetag{2.5} let's introduce the
following notations
$$
\varphi=\left<\tau, \bold N\right>\hskip 3em
\psi_i=\left<\tau, \bold M_i\right>\tag2.7
$$
Differentiating \thetag{2.7} and taking into account
\thetag{2.1}, \thetag{2.2}, \thetag{2.6} and \thetag{2.5} we get
the following equations
$$
\aligned
\dot\varphi &=w+\frac{B^i \psi_i}{v} \\
\dot\psi_i &= v G_{ik} z^k + \frac{B^k}{v}
\left(\vartheta^p_{ik} \psi_p - G_{ik} \varphi\right)
\endaligned\tag2.8
$$
For the space gradients of $A$ and $B^i$ we may define the
expansions
$$
\frac{\partial A}{\partial r^k} = a N_k+\alpha^p M_{pk}
\hskip 3em
\frac{\partial B^i}{\partial r^k}=b^i N_k + \beta^{ip} M_{pk}
\tag2.9
$$
Substituting \thetag{2.9} into \thetag{2.6} we derive the
following equations
$$
\aligned
\dot w &=a\varphi+\alpha^k\psi_k +
\frac{\partial A}{\partial v} w + \frac{\partial A}{\partial u^k}
z^k \\
\dot z^i &= -\frac{B^i w}{v^2} + \frac1v
\left(v b^i \varphi + \beta^{ik} \psi_k +
 \frac{\partial B^i}{\partial v} w +
 \frac{\partial B^i}{\partial u^k} z^k \right)
\endaligned
\tag2.10
$$
The equations \thetag{2.8} and \thetag{2.10} form the complete
system of linear differential equations with respect to
$\varphi$, $\psi_i$, $w$, $z_i$. Let's differentiate first of
the equations \thetag{2.8} by $t$ keeping in mind all above
expressions for the time derivatives of the quantities
involved. As a result we obtain
$$
\align
\ddot\varphi &= \left(a - \frac{B^q B^k}{v^2} G_{qk} \right)
\varphi + \left(\frac{\partial A}{\partial v} \right) w +
\left(\frac{\partial A}{\partial u^i} + G_{ik} B^k \right)
z^i + \\
&+ \left(\alpha^i + \frac{B^q B^k}{v^2} \vartheta^i_{qk} -
\frac{B^i A}{v^2} + b^i + \frac1v \frac{\partial B^i}{\partial v}
A + \frac{\partial B^i}{\partial u^k}\frac{B^k}{v^2} \right)
\psi_i
\endalign
$$
Let's consider the following expression denoted by $L$
$$
\ddot\varphi - P \dot\varphi - Q \varphi = L \tag2.11
$$
with $P$ and $Q$ being the coefficients enclosed in brackets in
the above expression for $\ddot\varphi$
$$
P = \left(\frac{\partial A}{\partial v}\right) \hskip 3em
Q = \left(a - \frac{B^q B^k}{v^2} G_{qk}\right)
$$
For $L$ in \thetag{2.11} then we derive the following expression
also being the linear combination of the expressions
``in brackets''
$$
L = \left(\frac{\partial A}{\partial u^i} + G_{ik} B^k \right)
z^i + \left(\alpha^i + \frac{B^q B^k}{v^2} \vartheta^i_{qk} -
\frac{B^i A}{v^2} + b^i + \frac{A}{v}
\frac{\partial B^i}{\partial v} - \frac{B^i}{v}
\frac{\partial A}{\partial v} \right) \psi_i
$$
Since $z^i$ and $\psi_i$ form the linearly independent set of
functions $L$ can vanish if and only if these ``brackets''
vanish. This gives us the following equations for $A$ and
$B^i$
$$
\align
& B^i = - G^{ik} \frac{\partial A}{\partial u^k}\tag2.12 \\
& {\aligned \alpha^i &+ \frac{B^q B^k}{v^2}\vartheta^i_{qk} -
\frac{B^i A}{v^2} + b^i + \\
&+ \frac{A}{v} \frac{\partial B^i}{\partial v} +
\frac{\partial B^i}{\partial u^k}\frac{B^k}{v^2} -
\frac{B^i}{v}\frac{\partial A}{\partial v} = 0 \endaligned}
\tag2.13
\endalign
$$
being the generalizations of \thetag{3.27} and \thetag{3.28}
from \cite{1} for the multidimensional case. \par
\definition{Definition 3} The dynamical system \thetag{1.1} with
force field of the form \thetag{2.3} is called the system with
the weak normality condition if the equations \thetag{2.12} and
\thetag{2.13} hold.
\enddefinition
     For the systems with the weak normality function $\varphi$
satisfies the second order ordinary differential equation derived
from \thetag{2.11}\par
$$
\ddot\varphi - P \dot\varphi - Q \varphi = 0 \tag2.14
$$
     For the dynamical system  of definition 3 to be  the system
accepting the normal shift in the sense of definition 2 we should
be able to obtain the initial conditions
$$
\left.\varphi\right|_{t=0} = 0 \hskip 3em
\left.\dot\varphi\right|_{t=0} = 0
$$
for any submanifold of codimension 1 by the choice of the modulus
of initial velocity $v=|\bold v|$. Let $S$ be some arbitrary
manifold of codimension 1 in $\Bbb R^n$. Defining the unit normal
vector for each point on $S$ we define the spherical map
$S\longrightarrow S^{n-1}$ from $S$ to unit sphere $S^{n-1}$ in
$\Bbb R^n$. For the submanifolds of general position this map is
the local diffeomorphism. The latter fact let us transfer the
coordinates $u^1,\dots,u^{n-1}$ (see above) from the unit sphere
$S^{n-1}$ to $S$ vector $\bold N(u^1,\dots,u^{n-1})$ from
\thetag{2.1} being the common unit normal vector for both.
Tangent vectors to $S$ are defined like $\bold M_i$ in
\thetag{2.1}
$$
\bold E_i=\frac{\partial \bold r}{\partial u^i} \hskip 3em
\frac{\partial\bold E_i}{\partial u^j} = \Gamma^k_{ij}
\bold E_k + b_{ij} \bold N \tag2.15
$$
Tensor $b_{ij}$ in \thetag{2.15} is the second quadratic form
for $S$ and $\Gamma^k_{ij}$ are the components of metric
connection on $S$  while $\vartheta^k_{ij}$ form the metric
connection on $S^{n-1}$. From \thetag{2.1} and \thetag{2.15}
we also derive
$$
\bold M_i = - b^k_i \bold E_k \hskip 3em
G_{ij} = b^k_i b^q_j g_{kq} \tag2.16
$$
For $S$ of general position the matrix $b^k_i$ is nondegenerate.
Let $d=b^{-1}$ be the inverse matrix. Then from \thetag{2.16} we
have
$$
\bold E_i = - d^k_i \bold M_k \hskip 3em
g_{ij} = d^k_i d^q_j G_{kq} \tag2.17
$$
Components of matrix $d^k_i$ in \thetag{2.17} form the
$G$-symmetric tensor
$$
d^k_i G_{kj} = G_{ik} d^k_j = b_{ij} \hskip 3em
\nabla_ib_{jk} = \nabla_jb_{ik} \tag2.18
$$
Second of the equations \thetag{2.18} known as the
Peterson-Coddazy equation holds with respect to both connections
$\Gamma^k_{ij}$ and $\vartheta^k_{ij}$. It is known that the
difference of two connections is the tensor
$$
Y^k_{ij} = \Gamma^k_{ij} - \vartheta^k_{ij} = b^k_q
\nabla_i d^q_j \tag2.19
$$
Covariant derivatives in \thetag{2.18}, \thetag{2.19} and
everywhere below are defined by the spherical connection
$\vartheta^k_{ij}$ on the unit sphere $S^{n-1}$. \par
     According to the definition 2 we are to determine the scalar
function $v=f(u^1,\dots,u^{n-1})$ such that
$$
\left.\dot\bold r\right|_{t=0} = f(u^1,\dots,u^{n-1})
\bold N(u^1,\dots,u^{n-1}) \tag2.20
$$
(compare with \thetag{3.23} from \cite{1}). Variables
$u^1=u^1(0),\dots,u^{n-1}=u^{n-1}(0)$ here play the same role as
$s$ in \thetag{3.23} from \cite{1}. Denoting $s=u^i(0)$ for a
while from \thetag{2.5}, \thetag{2.7} and \thetag{2.20} we obtain
$$
\left.\varphi\right|_{t=0} =
\left.\left<\bold N,\partial_s \bold r\right>\right|_{t=0} =
\left<\bold N,\bold E_i\right> \equiv 0
$$
So first initial condition for the equation \thetag{2.14} is
identically zero since all particles are starting from $S$
along the normal vector to this submanifold. For the second we
have
$$
\left.\dot\varphi\right|_{t=0} =
\left.\left<\partial_t\bold N,\partial_s \bold r\right>
\right|_{t=0}+
\left.\left<\bold N, \partial_{st} \bold r\right>\right|_{t=0}
$$
Using \thetag{2.20} and the latter expression we find the
following one
$$
\left.\dot\varphi\right|_{t=0} =
- \frac{b_{ik}}{f} B^k + \frac{\partial f}{\partial u^i}
\tag2.21
$$
To make zero  the initial condition \thetag{2.21} we need to
choose the function $f$ satisfying the following equations
$$
\frac{\partial f}{\partial u^i} =
\frac{b_{ik}(u^1,\dots,u^{n-1})B^k(\bold r,f,u^1,\dots,u^{n-1})}
{f}\tag2.22
$$
where $\bold r=\bold r(u^1,\dots,u^{n-1})$ is the vector of
cartesian coordinates of a point on $S$. Equations \thetag{2.22}
are analogs of \thetag{3.25} from \cite{1}. Since the equations
\thetag{5.22} form the overdetermined system of differential
equations one can derive from them some other equations being
the compatibility conditions for \thetag{2.22}. The way of
obtaining them is standard: one should differentiate
\thetag{2.22} by $u_j$ and then use the equality $\partial_{ij}f=
\partial_{ji}f$ by changing the order of derivatives. As a result
of such calculations we get
$$
\aligned
& \frac1v \frac{\partial B^k}{\partial v} B^q - \beta^{kq} =
  \frac1v \frac{\partial B^q}{\partial v} B^k - \beta^{qk} \\
& \nabla_kB^q = \frac{\nabla_pB^p}{n-1} \delta^q_k
\endaligned \tag2.23
$$
Functions $B^k$ in \thetag{2.23} are considered as the functions
of $2n$ independant variables $r^1,\dots,r^n$, $v$ and
$u^1,\dots,u^{n-1}$ as in \thetag{2.3}. Covariant derivatives by
$u^1,\dots,u^{n-1}$ are respective to the spherical connection
$\vartheta^k_{ij}$. \par
\proclaim{Theorem 2} The equations \thetag{2.12}, \thetag{2.13}
and \thetag{2.23} form the enough condition for the dynamical
system \thetag{1.1} with force field \thetag{2.3} to be
accepting the normal shift as described by the definition 2.
\endproclaim
     Equations \thetag{2.12}, \thetag{5.13} and \thetag{5.23}
are compatible in some sense since they have common solution for
$A$ and $B^k$ (at least trivial one with $A=A(v)$ and
$B^k\equiv 0$ corresponding to the geometrical situation from
\cite{5} and \cite{6}). Detailed analysis of these equations and
nontrivial examples of the multidimensional dynamical systems
associated with their solutions are the subject of separate
paper. \par
      Authors are grateful to Russian Fund for Fundamental
Researches for the financial support (project \# 93-011-273).
\par

\enddocument